\documentclass[aps,showpacs]{revtex4}
\usepackage{amssymb}
\usepackage[dvips]{graphicx}
\usepackage[english]{babel}
\usepackage{indentfirst}
\usepackage{amsxtra}
\usepackage{amsmath}
\usepackage{supertabular}
\usepackage{multirow}
\usepackage [mathcal]{eucal}
\def\beq{\begin{equation}}
\def\eeq{\end{equation}}
\def\beqn{ \begin{eqnarray} }
\def\bar{ \begin{array} }
\def\bdm{ \begin{displaymath}}
\def\esl{ \end{slide}}
\def\eeqn{ \end{eqnarray} }
\def\ear{ \end{array} } 
\def\edm{ \end{displaymath}}
\def\s1s2{{ \boldsymbol{\sigma}(1) \cdot \boldsymbol{\sigma}(2) }}
\def\t1t2{{ \boldsymbol{\tau}(1) \cdot \boldsymbol{\tau}(2)  }}

\def\ap{a_p}
\def\ap'{a_p'}
\def\ap1{a_{p_1}}
\def\ap'1{a_{p'_1}}
\def\ap2{a_{p_2}}
\def\ap'2{a_{p'_2}}
\def\ah{a_h}
\def\ah'{a_h'}
\def\ah1{a_{h_1}}
\def\ah'1{a_{h'_1}}
\def\ah2{a_{h_2}}
\def\ah'1{a_{h'_2}}
\def\acp{a\dag_p}
\def\acp'{a\dag_p'}
\def\acp1{a\dag_{p_1}}
\def\acp'1{a\dag_{p'_1}}
\def\acp2{a\dag_{p_2}}
\def\acp'2{a\dag_{p'_2}}
\def\ach{a\dag_h}
\def\ach'{a\dag_h'}
\def\ach1{a\dag_{h_1}}
\def\ach'1{a\dag_{h'_1}}
\def\ach2{a\dag_{h_2}}
\def\ach'1{a\dag_{h'_2}}

\newcommand{\bsigma}{\mbox{\boldmath $\sigma$}}
\newcommand{\btau}{\mbox{\boldmath $\tau$}}

\newcommand{\car}{$^{12}$C~}
\newcommand{\oxy}{$^{16}$O~}
\newcommand{\caI}{$^{40}$Ca~}
\newcommand{\caII}{$^{48}$Ca~}

\newcommand{\pb}{$^{208}$Pb~}
\begin{document}

\noindent
\title{Low-lying magnetic excitations of doubly-closed-shell nuclei
and nucleon-nucleon effective interactions}


\author{V. De Donno, G. Co' and C. Maieron}
\affiliation{Dipartimento di Fisica, Universit\`a del Salento and,
 INFN Sezione di Lecce, Via Arnesano, I-73100 Lecce, ITALY}
\author{M. Anguiano, A.M. Lallena and M. Moreno Torres }
\affiliation{Departamento de F\'\i sica At\'omica, Molecular y
  Nuclear, Universidad de Granada, E-18071 Granada, SPAIN}

\date{\today}

\begin{abstract}
  We have studied the low lying magnetic spectra of $^{12}$C, $^{16}$O,
  $^{40}$Ca, $^{48}$Ca and $^{208}$Pb nuclei within the Random Phase
  Approximation (RPA) theory, finding that the description 
  of low-lying magnetic
  states of doubly-closed-shell nuclei imposes severe constraints on
  the spin and tensor terms of the  nucleon-nucleon effective 
  interaction. We have first made an investigation by using four
  phenomenological effective interactions and we have obtained good
  agreement with the experimental magnetic spectra, and, to a lesser
  extent,  with the
  electron scattering responses. Then we have made self-consistent RPA
  calculations to test the validity of the finite-range D1 Gogny
  interaction. For all the nuclei under study we have found that this
  interaction inverts the energies of all the magnetic states forming
  isospin doublets.
\end{abstract}

\maketitle

\section{Introduction}
\label{sec:intro}
In the last thirty years, electron scattering experiments on nuclei
have produced an enormous amount of high-precision, accurate and
reliable data which impose severe constraints on nuclear models and
theories. 

Our interest is focused on the excitation of
unnatural parity states in the low-lying region of the nuclear
spectrum, where many responses of several nuclei have been measured
\cite{hei83,don68,lic79,hic84,mul85,wis85,but86,hic87,hyd87,con92}.
The description of these  states with effective
theories, such as the Random Phase Approximation (RPA), indicates a
strong sensitivity to the details of the spin and tensor dependent
terms of the  Nucleon-Nucleon (NN) effective interactions.  While the
study of a single excited state, or of a limited set of excited
states, for a single nucleus has been pursued in depth, as for
example in Refs.  \cite{co90,lal93,lal96}, a systematic study of a
large set of nuclei and of excited states has not been presented, and
the availability of many precise experimental data has not been fully
exploited.

We present here results of such a systematic study, which indicate
that there are general requirements that the NN effective interaction
has to fulfill, in order to provide a reasonable description of the
low-lying magnetic excitations. We have obtained these results by
using a phenomenological approach to the RPA theory inspired to the
Landau-Migdal theory of finite Fermi systems \cite{mig67,spe77}.  In
this approach the Mean-Field (MF) basis, which provides the set of single
particle energies and wave functions to be used in the RPA
calculations, 
is generated by  a
Woods-Saxon well, whose parameters are adjusted to reproduce at best
some ground state properties of the nucleus, such as the charge
density distribution and the single particle energies around the Fermi
surface. In addition, a phenomenological residual NN effective 
interaction is used. The parameters of this interaction are chosen to
reproduce the energy of some specific excited states. In terms of
comparison with the experimental data, this phenomenological approach
uses the RPA theory at its best. 

In order to study the sensitivity of
our results to the details of the residual interaction, we have
developed four phenomenological interactions, two of them having
zero-range, as in the original formulation of the Landau-Midgal
theory, and the other two having finite-range. For each type of
interaction we have considered a parametrization which includes
tensor terms and another one without them, and we
have used these interactions to study the excitation
of the low-lying magnetic spectra of $^{12}$C, $^{16}$O, $^{40}$Ca,
$^{48}$Ca and $^{208}$Pb nuclei. We have found only few cases that are
sensitive to the differences between the various interactions, and we
present them in the paper.  The main result of our study is, however,
that most of the states are equally well described by all
the interactions we have considered. This suggests
that we have been able to include some general features of the
interaction, which are necessary for the description of
the magnetic excitation spectra
of doubly-closed-shell nuclei.

In order to test this hypothesis we
have then repeated our RPA calculations 
within a fully self-consistent approach.
This means that the MF states and energies are obtained within the 
Hartree-Fock (HF) approximation, using the same NN effective interaction
employed in the RPA calculations. In particular we have used the
Gogny D1 finite-range interaction \cite{gog75,bla77,dec80}. 
In this
case we have found remarkable disagreement with the experimental
data,
the most
striking result being that all the energies of the states which form an
isospin doublet are inverted.  This indicates that the good results
obtained with the phenomenological approach are not accidental, and
that the study of the magnetic spectra is selective in choosing the
strength of the relevant terms of the force.

The paper is organized as follows. In Sec. \ref{sec:details} we
give some detail of our calculations, mainly regarding the input 
used to solve the RPA equations. 
The results of our phenomenological study of several magnetic states for
all the nuclei under investigation are presented in Sec.  \ref{sec:pheno}. In
Sec. \ref{sec:HF} we give the results of the self-consistent
calculation using the Gogny D1 interaction, for some selected
cases. Finally, in Sec. \ref{sec:conclu}, we draw our conclusions.

\section{Details of the calculation}
\label{sec:details}

The first input required by the RPA calculation is the set of single
particle wave functions and energies. In the phenomenological
calculations we have used the single particle bases generated by
Woods-Saxon wells. The parameters of the wells have been taken from
the literature \cite{ari07}, and have been chosen to describe at best
the energies of the single particle states around the Fermi surface
and the ground state charge density distributions. In the
self-consistent calculations, the single particle wave functions and
energies have been obtained by solving the Hartree-Fock equations with
the method described in Refs. \cite{co98b,bau99}.

We have solved the RPA equations by using a discrete set of single particle
wave functions and energies. In the phenomenological calculations, the
discretization of the continuum has been obtained by diagonalizing the
Woods-Saxon well in a harmonic oscillator basis. In the
self-consistent calculations the discretization has been obtained by
imposing  the correct boundary conditions of a bound state to the
single  particle wave functions at the edge of the computing box. 
The global RPA solutions strongly depend on the size of the single
particle configuration space \cite{don08t}. However, there are excited
states dominated by particle-hole excitations where the particle wave
function is bound.  In this article we consider only this type of
states. 

For each nucleus considered, we have used single particle
configuration spaces large enough to ensure the stability of the
results for the states under investigation. In the phenomenological
calculation the smallest configuration space, used for $^{12}$C, is
composed of 5 major harmonic oscillator shells, for a total of 44
single particle states. The largest space has been used for
$^{208}$Pb, and it is composed by 9 major shells for protons and 10
major shells for neutrons, for a total of 100 single particle states.
In the self-consistent calculations we have fixed the size of the
computational box, $R_{\rm max}$, and the maximum energy of the
particle states in the configuration space, $E_{\rm cut}$. In the case
of $^{12}$C, $R_{\rm max}=10$~fm and $E_{\rm cut}=50$~MeV, while for
$^{208}$Pb \/ these two parameters are 14~fm and 50~MeV, respectively.

The second input required by  RPA is the residual interaction,
which, in analogy to the microscopic nucleon-nucleon (NN) interactions
of Urbana or Argonne type, we write as 
\beqn
V_{\rm eff}(1,2)&=&
       v_1(r_{12}) \, 
  + \, v_1^\rho(r_{12})\, \rho^\alpha(r_1,r_2) \nonumber \\
&&+ \, \left[ v_2(r_{12}) \, 
  + \, v_2^\rho(r_{12}) \, \rho^\alpha(r_1,r_2) \right] \, \t1t2\nonumber \\
&&+ \, v_3(r_{12})\, \s1s2 \, 
  + \, v_4(r_{12}) \, \s1s2 \, \t1t2  \nonumber \\
&&+ \, v_5(r_{12}) \, S_{12}(\hat{r}_{12}) \, 
  + \, v_6(r_{12})\, S_{12}(\hat{r}_{12}) \, \t1t2 
\, .
\label{eq:intr}
\eeqn
Here, following the indications of past phenomenological \cite{spe77}
and self-consistent \cite{bla77} RPA studies, we have 
included a possible dependence on the nuclear one-body density $\rho(r)$
in  the central and isospin channels. 
In Eq.~(\ref{eq:intr})
$r_{12}=|\boldsymbol{r}_1-\boldsymbol{r}_2|$, $\bsigma$ and $\btau$
are the usual spin and isospin operators, $S_{12}$ is the tensor
operator defined as
\beq S_{12}(\hat{r})\, = \, 3 \,
\boldsymbol{\sigma}(1)\cdot\hat{r} \,
\boldsymbol{\sigma}(2)\cdot\hat{r} \, - \, \s1s2 
\label{eq:s12}
\eeq
and 
\beq
\rho(r_1,r_2) \, = \, {\left[\rho(r_1)\rho(r_2)\right]}^{1/2} 
\, .
\label{eq:vdens2}
\eeq

The $v_i(r)$ functions of Eq.~(\ref{eq:intr}) are the same for all
nuclei under investigation. On the other hand, the $v_i^\rho(r)$
corresponding to the density dependent part of the interaction are
assumed  to be different for each nucleus: they have been
chosen to reproduce the first $2^+$ state in $^{12}$C and the first
$3^-$ state in $^{16}$O, $^{40}$Ca, and $^{208}$Pb. The other terms
of the force have been chosen to get a reasonable description of the
centroid energy of the isovector giant dipole resonance by caring that
the isoscalar spurious $1^-$ excitation is at zero energy or below.
These criteria are useful for the scalar and isospin terms of the
interaction, the main responsible for the excitation of natural parity
states. The $v_i(r)$ functions of the spin, spin-isospin and tensor
channels of the interaction ($i=$3,4,5 and 6) have been adjusted
to describe the excitation energies of the magnetic states below 8 MeV
in $^{208}$Pb, paying particular attention to the $1^+$ states at
$5.85$ and $7.30$ MeV \cite{las85}, and to the $12^-$ states at $6.43$
and $7.08$ MeV \cite{lic79}. In addition, we have also cared that the
correct sequence of the two 1$^+$ states in $^{12}$C forming an
isospin doublet \cite{don68} is obtained, and that the 
energy of the first $4^-$
state of $^{16}$O \cite{hyd87} is reasonably reproduced.

In this work we are interested in the possible effects of the tensor
channels of the interaction as well as in the relevance of its range
(zero or finite). Thus, we have built four interactions. In
connection with previous RPA studies \cite{co90} we have considered two
interactions, based on the Landau-Migdal approach and
labeled LM and LMtt in the following, which have zero-range. 
For these two cases the functions $v_i(r)$ 
of Eq.~(\ref{eq:intr}) are given by
\begin{equation}
v_i(r_{12})\, = \, V_i \, \delta(\boldsymbol r_1 -\boldsymbol r_2)
\, , \,\,\,\,\, i=1,\ldots,6 \, .
\label{eq:LM}
\end{equation} 
The values of the parameters $V_i$, in MeV fm$^3$, are
\[
V_1\, = \, -918\,; \,\, V_2\,=\, 600\,; \,\, V_3 \, = \, 20\,;\,\,
V_4\,=\,200\,; \,\, V_5\,=\,0\, .
\]
For the LM interaction, $V_6=0$, while for the LMtt one, $V_6=-150$
MeV fm$^3$.

Also the terms $v_i^\rho(r)$ of Eq.~(\ref{eq:intr}) have zero-range 
\begin{equation}
v_i^\rho(r_{12})\, = \, V_i^\rho \, \delta(\boldsymbol r_1 -\boldsymbol r_2)
\, , \,\,\,\,\, i=1,2 \, .
\label{eq:LMrho}
\end{equation} 
In MeV fm$^6$ units, the values of $V_1^\rho$ are 361.0, 436.4, 492.3
and 599.0 and those of $V_2^\rho$ are -40.0, -31.0, -150.0 
and 0.0
for the $^{12}$C, $^{16}$O, $^{40}$Ca \/ and $^{208}$Pb
respectively. For $^{48}$Ca, we have used the same values as for
$^{40}$Ca. 
In all the calculations within the phenomenological approach we have used
$\alpha$=1 in Eq. (\ref{eq:intr}).

In our phenomenological RPA approach we have considered only the
contribution
of direct matrix elements, assuming
that the effects of the exchange terms are effectively included in the
choice of the parameters of the various interactions.
Therefore the scalar
and isospin terms, $v_1$ and $v_2$ respectively,
do not contribute to the excitation of unnatural parity states, which
are the focus of this work.  For
sake of completeness, however, we present here the full effective interactions.

We have also considered two finite-range interactions with and
without the tensor terms, which we have labeled FRtt and FR, respectively.
They are obtained from the Argonne V18
potential \cite{wir95}, by modifying its short range behavior to take
into account short range correlations effects. In particular, the
short range part of the Argonne V18 NN potential is removed and
replaced by a combination of Gaussian functions. Specifically, we have
taken
\beq
v_i(r)\, = \, \widetilde V_{18}^i(r) \,
+ \, \sum_{\mu=1}^M \, a^i_\mu \, \exp 
\left[ {-b^i_\mu\,(r-R^i_\mu)^2} \right]
\, , \,\,\,\,\, i=1,\ldots,4 \, ,
\label{eq:FR14}
\eeq
where $\widetilde V_{18}^i(r)$ are the corresponding terms of the bare
Argonne V18 potential with their short range terms set to zero. In
Eq.~(\ref{eq:FR14}) $M$ is the number of Gaussians used in each
channel.  For the scalar channel we have included two Gaussians in
order to obtain an attractive behavior starting from the repulsive
core. The repulsive behavior is accounted for by the density dependent
term. For the channels $i=2$ and $i=4$ we have used only one Gaussian
and
we have set to zero the spin term $i=3$.
The values of the various parameters are given in Table
\ref{tab:FRpar} and they are the same for both interactions.

\begin{table}[ht]
\begin{center}
\begin{tabular}{cccccccc}
\hline\hline
  &~~& $a_1^i$ & $b_1^i$ & $R_1^i$ & $a_2^i$ & $b_2^i$ & $R_2^i$ \\
channel &
& [MeV]  & [fm$^{-2}$] & [fm] & [MeV]  & [fm$^{-2}$] & [fm] \\
\hline
$i=1$ && 600.0 & 4.0 & 0.5 & -200.0 & 20.0 & 0.0 \\
$i=2$ && 300.0 & 7.0 & 0.5 &        &      &     \\
$i=4$ && -40.0 & 4.5 & 0.5 &        &      &     \\
\hline\hline
\end{tabular}
\end{center}
\caption{\small Parameters of the Gaussian functions
of the FR and FRtt interactions (see Eq. (\ref{eq:FR14})). 
The spin terms, $i=3$, have been set to zero.}
\label{tab:FRpar}
\end{table}

In the FRtt case the tensor channels have been obtained by
multiplying the bare V18 tensor terms by the scalar term of the
two-body short range correlation function $f(r)$ of Ref. \cite{ari07}
\beq
v_i(r) \, = \, V_{18}^i(r) \, f(r)
\, , \,\,\,\,\, i=5,6 \, .
\label{eq:FRtensor}
\eeq
More specifically, we have used the correlation functions obtained with the
so-called Euler procedure and, because of the small differences
between the $f(r)$ of the various nuclei \cite{ari07}, 
we have used the function obtained for the $^{40}$Ca in all our calculations. 
In the FR interaction the tensor terms are equal to 0.

Finally, the density dependent terms have been taken to be Gaussians:
\beq
v_i^\rho(r) \, = \, A_i \, \exp \left(- B_i \, r^2\right)
\, , \,\,\,\,\, i=1,2 \, .
\label{eq:FRrho}
\eeq
In our calculations we have used $B_i=1$fm$^{-2}$. 
The values of the parameter $A_i$ are shown in Table \ref{tab:FRrho}.

\begin{table}[ht]
\begin{center}
\begin{tabular}{ccccccc}
\hline\hline
&~~& \multicolumn{2}{c}{FR} &~~& \multicolumn{2}{c}{FRtt} \\ 
\cline{3-4} \cline{6-7}
   && $A_1$ & $A_2$ & & $A_1$ & $A_2$ \\
nucleus     && [MeV]    & [MeV]   & & [MeV]    & [MeV]       \\
\hline
$^{12}$C    && 133.8    & -120.0  & & 133.8    & -125.0   \\
$^{16}$O    && 163.4    &  -95.0  & & 163.6    &  -95.0   \\
$^{40}$Ca   && 194.7    &  -50.0  & & 194.6    &  -50.0   \\
$^{208}$Pb  && 240.0    &  -25.0  & & 240.0    &  -25.0   \\
\hline
\hline
\end{tabular}
\end{center}
\caption{\small Parameters of the density dependent terms of the FR
and FRtt interactions (see Eq. (\ref{eq:FRrho})).
}
\label{tab:FRrho}
\end{table}

The choice of the free parameters of the finite-range interactions has
been done following the same criteria used for the zero-range
interactions, and also in this case
we have included in the RPA calculations only the contributions of the
direct terms of the matrix
elements.

In our self-consistent RPA calculations we have employed a 
Gogny interaction \cite{gog75,bla77,dec80} which is usually expressed as 
\beqn
V_{\rm eff}(1,2) \, = && \sum_{i=1}^2 \,
\exp \left[
-\frac{\left( \boldsymbol{r}_1-\boldsymbol{r}_2 \right )^2}{\mu_i^2} 
\right] \,
(W_i\, +\,B_i \, \hat P_\sigma \, - \, H_i \, \hat
P_\tau \, - \, M_i \, \hat P_\sigma \, \hat P_\tau) \nonumber\\
&&+ \, W_{LS} \, \left( 
\boldsymbol{\sigma}(1)\, + \, \boldsymbol{\sigma}(2) \right)
\,\,\,
^{^{^{\leftarrow}}} \!\!\!\!k
\, \times \, \delta(\boldsymbol {r}_1-\boldsymbol
{r}_2) \, 
\vec{k}
\nonumber\\
&&+ \, t_0 \, \left( 1\, +\, x_0 \, \hat P_\sigma \right) \, 
\delta(\boldsymbol {r}_1-\boldsymbol
{r}_2) \, \rho^\alpha \left( \frac{1}{2}(\boldsymbol{r}_1\, +\, \boldsymbol
{r}_2)\right)
\, ,
\label{eq:gogny_int}
\eeqn
where $\vec{k}$ is the operator of the relative momentum 
\beq
\vec{k} \, = \, \frac{1}{2i} \, \left( 
\boldsymbol{\nabla}_1 \, -  \boldsymbol{\nabla}_2 \right) 
\,\,\,.
\eeq
We have indicated with $\hat P_{\sigma}$ and $\hat P_{\tau}$ the
usual spin and isospin exchange operators, and $\mu_i$, $W_i$, $B_i$,
$H_i$, $M_i$, $W_{LS}$, $t_0$ and $x_0$ are constant parameters.

The relation between the expression above of the Gogny force and that
required by Eq. (\ref{eq:intr}) is obtained from the following equations:
\begin{eqnarray}
v_1(r) &=& W(r) \, +\, \frac{B(r)}{2}\, 
-\,\frac{H(r)}{2}\, -\, \frac{M(r)}{4} \, ,\\
v_2(r) &=& \frac{B(r)}{2}\, -\, \frac{M(r)}{4} \, , \\
v_3(r) &=& -\frac{H(r)}{2}\, -\, \frac{M(r)}{4} \, , \\
v_4(r) &=& - \frac{M(r)}{4} 
\,\,,
\end{eqnarray}
where
\beq
F(r) \, = \, \sum_{i=1}^2 \,
 F_i \, \exp \left(
-\frac{r^2}{\mu_i^2} 
\right) \, , \,\, \,\,\, F \equiv W, B, H, M \, . 
\eeq

The density dependent term of Eq.~(\ref{eq:gogny_int}) can
be written as:
\beq
t_0\,(1\, +\, x_0 \, \hat P_\sigma)\, \rho^\alpha \, 
= \, 
\left[ t_0 \, \left(1\,-\, \frac{x_0}{2} \right) \, - \, 
\frac{t_0\, x_0}{2}\, \t1t2 \right]\, \rho^\alpha  \, .
\eeq

In particular  we have used the parametrization of the Gogny
interaction known as D1 \cite{gog75,bla77,dec80}. In the HF
calculations we have included all the terms of the interactions,
while in the RPA calculations we have neglected the contribution of
the spin-orbit term. In HF and RPA calculations both direct and
exchange terms of the interaction matrix elements have been
considered. 

\begin{figure}[ht]
\begin{center}
\includegraphics[scale=0.5, angle=90] {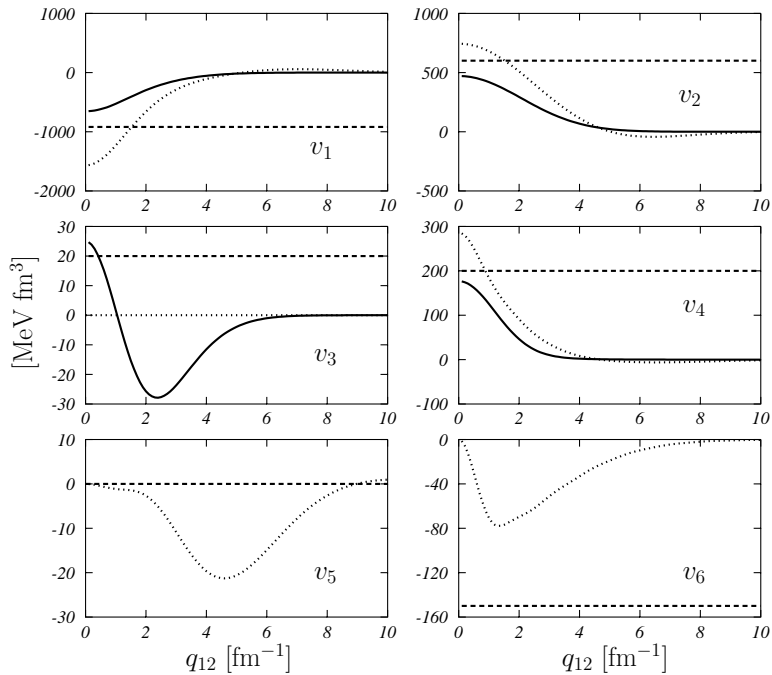} 
\caption{\small The effective 
 NN interactions used in this work as a function
 of the relative momentum. The solid lines represent the D1, 
the dashed lines the LMtt and 
dotted lines the FRtt interactions respectively. The central channels 
$v_1...v_4$ of LMtt and FRtt interactions
are identical to those of the LM and FR interactions respectively. 
}
\label{fig:force}
\end{center}
\end{figure}

The various interactions used in our work are shown in
Fig. \ref{fig:force} as a function of the relative momentum of the
interacting pair of nucleons.  In this figure, solid, dashed and
dotted lines represent the D1, LMtt and FRtt interactions,
respectively. Our tensor dependent interactions have been obtained
by adding the two tensor
dependent terms $v_{5,6}(r)$ to the LM and FR four central channels . 
For this reason, in the figure, the LM and FR
interactions are not shown, since they are identical, in the central
channels, to the LMtt and FRtt interactions. The zero-range
interaction terms are constant in momentum space. We point out that
the spin term $v_3$ has been set to zero in the FR and FRtt
interactions, and that LMtt does not have the pure tensor term
$v_5$. The finite-range interactions FR, FRtt and D1 have similar
asymptotic behavior, above 4 fm$^{-1}$. The values of the LM and LMtt
interactions fall between those of the D1, FR and FRtt
interaction at $q_{12} \sim 0$.

When a discrete configuration space of single particle
wave functions is used, the solution of the RPA equations is obtained
by
solving
a homogeneous system of linear equations. For a given excitation
multipole of angular momentum $J$ and parity
$\pi$, the RPA solution, obtained with standard diagonalization
procedures, provides the set of excitation energies, and, for each
excited state, the full set of RPA amplitudes $X^{J^\pi}_{ph}$ and
$Y^{J^\pi}_{ph}$. One can then calculate the amplitudes for the
transition between ground and excited states induced by an operator
$T_J(q)$ as
\beq
\langle J \| T_J(q) \| 0 \rangle \, = \, 
\sum_{p h}\,  
\left[ X^{J^\pi}_{p h} \, \langle j_p \| T_J(q) \| j_h \rangle \, + \, 
(-1)^{J+j_p -j_h}\, Y^{J^\pi}_{p h} \, 
\langle j_h \| T_J(q) \| j_p \rangle \right] 
\label{eq:transs1}
\, .
\eeq
In the equation above $|j \rangle$ indicates the single particle wave
function characterized by the set of quantum numbers including,
principal quantum number, orbital angular momentum, total angular
momentum $j$, and isospin third component. The double bars indicate
the reduced matrix elements of the angular coordinates.

In this work we have calculated  electromagnetic responses, which are
defined as the Fourier transform of the squared moduli of the
transition amplitude (\ref{eq:transs1}) \cite{hei83}. In the plane
wave Born approximation description of inelastic electron
scattering experiments, these responses, which depend on the modulus
of the momentum transfer $q$, are related to the cross section by
multiplicative factors depending on kinematics variables, and to the
Mott cross section \cite{for66,bof96}.

Since we are interested in magnetic states, the charge operator does
not contribute. The operators we have used to calculate the transition
amplitudes (\ref{eq:transs1}) are those of the convection and
magnetization currents. The explicit expressions of the single
particle matrix element can be found in Refs. \cite{co85,ama93}. We
have not considered the contribution of meson-exchange currents,
which, for low-lying excited states, has been found to be
negligible in comparison with the effects of the residual interaction
\cite{deh85,co90}.

\section{Results of the phenomenological approach}
\label{sec:pheno}

In this section we present our results for the low-lying magnetic
states of $^{12}$C, $^{16}$O, $^{40}$Ca, $^{48}$Ca and $^{208}$Pb,
obtained within the phenomenological approach.  For each nucleus we
first present the unnatural parity low energy spectrum, we compare it
with the measured spectrum, and we discuss the sensitivity of the
excitation energies to the inclusion of finite-range and tensor terms
contributions in the residual interaction.  Then, for some specific
states, we investigate the electromagnetic transverse response
functions.  In order to minimize the uncertainties due to the
discretization of the continuum, we have selected excited states which
are dominated by particle-hole (ph) pairs where the particle is in a
bound state. Furthermore, we have chosen the states which exhibit
the largest sensitivity to those terms of the residual interaction
which are the focus of the present study.  Also, we have addressed our
attention to those states forming isospin doublets, because their
structure (order of the states and relative splitting) is sensitive to
the isospin dependent terms of the residual interactions and, more
specifically, to the tensor-isospin terms we have introduced in the
previous section.  The interest in isospin doublets will become
clearer in connection to the self-consistent calculations which will
be presented in next section.  In addition, we give preference to
the study of those states for which experimental data are available.

A detailed discussion of the results will be presented throughout this
section, but we would like to anticipate that we have obtained a
general good description of the excitation spectra, almost
independently of the effective interaction used. This indicates that
we have been able to incorporate in the parametrization of the
residual interaction some relevant features required by the
description of the magnetic excitations. The disagreement with the
experimental data can be due to the use of the Plane Wave Born
Approximation in the calculation of the electron scattering cross
section, or in the nuclear structure part, to the truncation of the
configuration space. Actually these approximations are rather well
controlled. The experimental responses are usually presented after a
correction for the Coulomb distortion of the electron wave functions,
and the effects of the limited configuration space are effectively
considered by the choice of the force parameters. For these reason we
think that the possible discrepancies between our predictions and the
experimental data have to be ascribed more to the intrinsic
limitations of the RPA theory rather than to a more efficient
parametrization of the interaction.

\subsection{The $^{12}$C nucleus}
\label{sec:c12}

In Table \ref{tab:exc12} we compare the energies of the low-lying
magnetic states of $^{12}$C with the experimental values taken from
Ref. \cite{led78}. In the calculation with the LM interaction we have
been unable to identify the second 2$^-$ state, because all the states higher
than the first one have dominant ph components with the particle in the
continuum. Apart from this case we notice that the calculated energies
for each state are rather similar, independently on the interaction
used.  The experimental energies are reasonably well reproduced except
for the 2$^-$ states whose energies are about 4 MeV above the
experimental ones.

\begin{table}[ht]
\begin{center}
\begin{tabular}{cccccc}
\hline\hline
\multicolumn{6}{c}{$^{12}$C}\\
\hline\hline
$J^\pi$&LM&LMtt&FR&FRtt& exp \\
\hline
$2^-$ &  16.26 & 16.20 & 16.07 & 16.03 & 11.83\\
$1^+$ &  14.41 & 14.41 & 13.89 & 13.87 & 12.71\\
$2^-$ & ---    & 17.26 & 17.23 & 17.14 & 13.35\\
$1^+$ & 18.13  & 17.97 & 18.17 & 18.05 & 15.11\\
$4^-$ & 18.21  & 18.21 & 17.78 & 17.75 & 18.27\\
$4^-$ & 21.70  & 20.80 & 19.92 & 19.49 & 19.50\\
\hline\hline
\end{tabular}
\caption{\small Low-lying spectrum of the unnatural parity, magnetic,
  states in $^{12}$C. The energies are expressed in MeV. 
  The experimental values are from Ref. \cite{led78}.
\label{tab:exc12}
}
\end{center}
\end{table}

\begin{figure}[ht]
\begin{center}
\includegraphics[scale=0.5, angle=0] {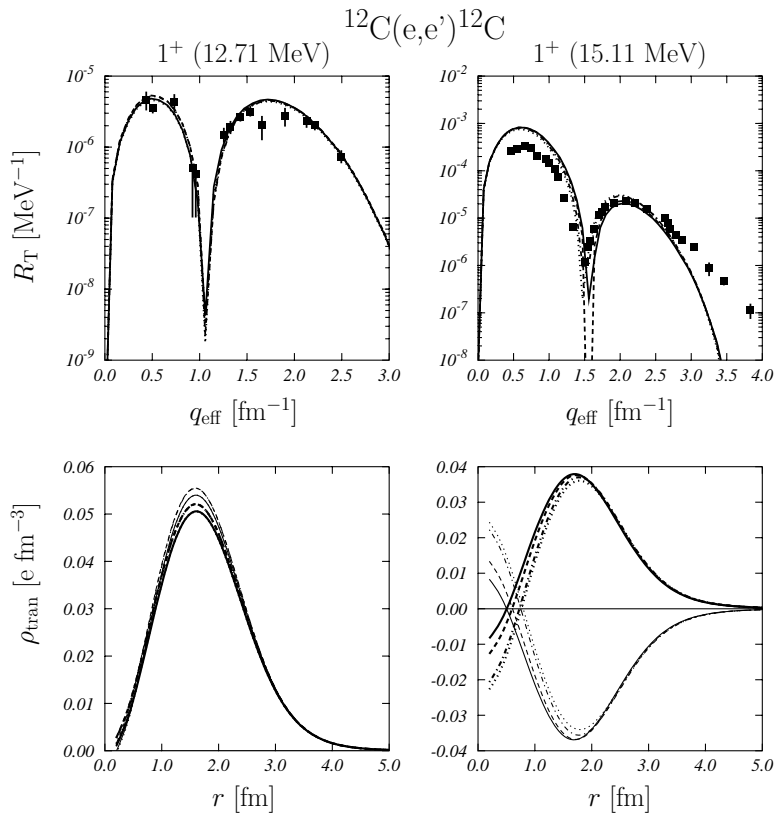} 
\caption{\small The Upper panels show the electromagnetic responses of
  the 1$^+$ states in \car. The data are taken from \cite{hyd87}. The
  various lines indicate the interaction used in the RPA calculation,
  specifically, LM (solid), LMtt (dotted), FR (dashed) and FRtt
  (dashed-dotted). 
  The lower panels show protons (thick lines) and neutrons
  (thin lines) contributions to the transition densities of the two
  states. The line types have the same meaning as in the upper panels.
  The values of the experimental energies for the two states are
  indicated.  
}
\label{fig:12C1+}
\end{center}
\end{figure}

The two most interesting cases are the $1^+$ and $4^-$ states.  For
the $1^+$ case we obtain two states, dominated by the 
$[(1p_{1/2}) (1p_{3/2})^{-1}]$ proton and neutron pairs, although for
the state with higher energy not negligible contributions of other ph
pairs appear.  The lowest energy state has isoscalar (IS) character
while the state with higher energy is isovector (IV). These
states correspond to the experimentally well known isospin doublet at
12.71 MeV (T=0) and 15.11 MeV (T=1) \cite{led78,hic87}. The
corresponding transverse response functions, or form factors, are
shown as a function of the effective momentum transfer in the upper
panels of Fig. \ref{fig:12C1+}. We use the traditional definition of
the effective momentum \cite{hei83}
\beq
q_{\rm eff} \, = \, q \, \left( 1 \, + \, 
\frac {3Z\alpha \hbar c}{2 \epsilon_i R} \right) \, ,
\eeq
where $Z$ is the atomic number of the target nucleus, $\alpha$ is the
fine structure constant, $\epsilon_i$ is the incident electron energy
and $R$ is the nuclear charge radius.

In the lower panels of the same figure, for each state considered, we
show the proton (thick lines) and neutron (thin lines) contributions
to the transition densities, as a function of the distance from the
center of the nucleus. These transition densities have been obtained
from Eq. (\ref{eq:transs1}) by considering for $T_J$ the expression
of the magnetization and avoiding the integration on $r$.  The
behavior of the transition densities clearly shows the isospin nature
of the two states. For the lower energy state, proton and neutron
densities are in phase, indicating the IS nature of the
excitation. The opposite happens for the second state and this is a
clear signature of the IV nature of this state.

As we can see from Table \ref{tab:exc12}, our calculations
overestimate the experimental energies of both the $1^+$ states and
also their splitting. The largest relative differences in the energies
values are 13\% and 20\% for the first and second state,
respectively. Despite these quantitative discrepancies with the
observed energies, our calculations produce the correct sequence of
isoscalar and isovector excitation with all the interactions. The
inclusion of finite-range and tensor terms changes the energy values
at the level of few percent.  Also the response functions are not very
sensitive to the use of different residual interactions as it is shown
in Fig. \ref{fig:12C1+}. Only the responses of the IV state show some
difference around the minimum at $q_{eff}$=1.5 fm$^{-1}$. The position
of this minimum seems to be slightly better described by the
interactions including tensor terms.
The comparison with electron scattering data \cite{hyd87} shows good
agreement with the IS data and overestimates the experimental IV
response in the region of the first maximum.  A good description of
the 1$^+$ IV transition is extremely important since this state is
used in liquid scintillator neutrino detectors to identify neutral
current events \cite{aga07}. The figure shows that the discrepancy in
the description of the IV response cannot be solved by using an
overall quenching factor. While the first peak is overestimated by
almost a factor two, the second peak is rather well reproduced.  The
difficulty in describing the IV 1$^+$ state is a common characteristic
of the RPA calculations \cite{kol94,kol95,vol00,co06b,co09a}, and
produces an overestimation of the experimental total neutrino $^{12}$C
cross sections measured in the LNSD \cite{ath98,agu01,aga07} and
KARMEN \cite{eit99} experiments.  In order to solve the problem, the
presence of strong pairing effects has been advocated \cite{krm05},
with the idea that the shell closure in the $^{12}$C nucleus is not a
good approximation. We have to remark, however, that the problem of
describing the the IV 1$^+$ state is present also in other doubly
magic nuclei where pairing correlations are negligible \cite{las85}.
The size of the first maximum of the 1$^+$ IV response in $^{12}$C is
well reproduced by microscopic {\sl ab initio} shell model
calculations \cite{hay03}, but the shape is completely wrong. These
calculations produce the first minimum of the response at 2 fm$^{-1}$,
and they are completely missing both size and shape of the second
maximum.

We consider now the $4^-$ states which also form an isospin doublet.
These states are dominated by the linear combination of the 
stretched $[(1d_{5/2}) (1p_{3/2})^{-1}]$ excitations.  In our
calculations the $1d_{5/2}$ state is bound  with an
energy of -1.1 MeV in the neutron case, 
and it shows a sharp resonance at 2.0 MeV in the
case of protons. The MF excitation energies are the single particle
energy differences, which for this ph transitions are 17.96 and 17.62
MeV for protons and neutrons, respectively. The RPA calculations mix
the proton and neutron ph transitions and in our results
the isoscalar state has lower
energy than the isovector one, independently of the interaction used.
The results shown in Table
\ref{tab:exc12} indicate that the residual interaction produces 
solutions with energies higher than those obtained  within the simple MF. In
this situation the role of the finite-range of the force is not
negligible. The upward shift of the RPA solutions is reduced by only
0.5 of MeV for the IS state, but by 1.7 MeV for the IV state. The
experimental IS energy is better reproduced by the zero-range
interaction, while the IV energy is much better described by the FRtt
interaction.

\begin{figure}[ht]
\begin{center}
\includegraphics[scale=0.5, angle=0] {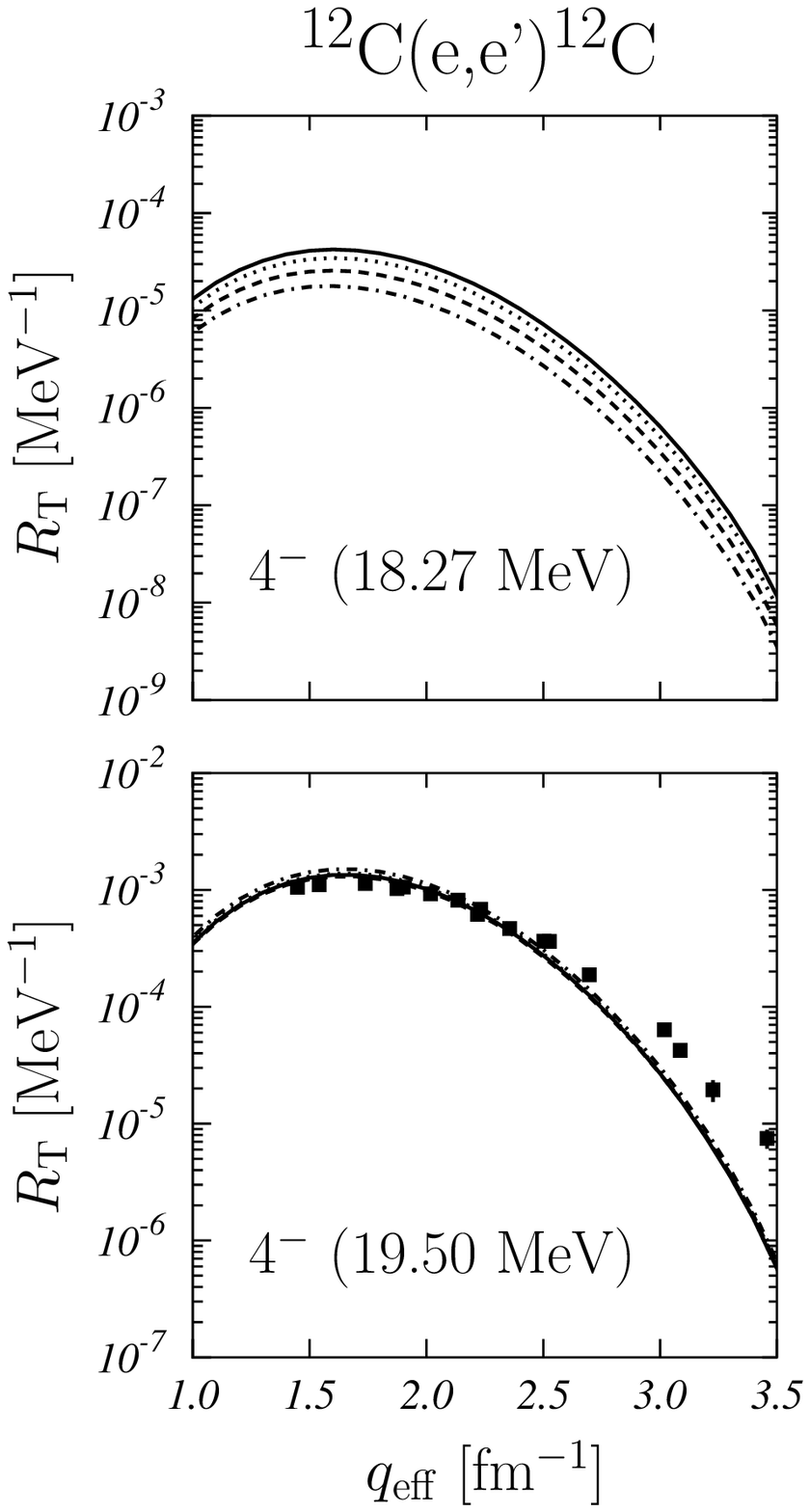} 
\caption{\small Electromagnetic responses of the 4$^-$ states in
  \car. The meaning of the lines is the same as in
  Fig. \ref{fig:12C1+}. The data are taken from \cite{hic84}.
  }
\label{fig:12C4-}
\end{center}
\end{figure}

We show in Fig. \ref{fig:12C4-}, the electromagnetic responses for the
two $4^-$ states. We compare the IV responses to the available
experimental data \cite{hic84}. The IS responses show some sensitivity
to the use of the residual interaction. The inclusion of the tensor
terms and of the finite-range, reduces the size of the response. The
results of the IV responses are rather independent of the residual
interaction and the experimental data are rather well reproduced. 

Finally we observe that our model also produces $2^-$ states, however,
as said before, their energies are in large disagreement with
data. The same occurs when the corresponding responses are
compared. This might be due to the presence, in these states, of
sizable contributions from ph pairs having a particle in the
continuum, which bring in further uncertainties, as our procedure
discretizes the continuum.  

\begin{table}[h]
\begin{center}
\begin{tabular}{cccccc}
\hline\hline
\multicolumn{6}{c}{$^{16}$O}\\
\hline\hline
$J^\pi$&LM&LMtt&FR&FRtt&exp \\
\hline
$2^-$ & $11.80$&$11.80$&$11.51$&$11.51$&8.87\\
$0^-$& 12.33& 11.19& 12.15 & 11.84 & 10.96\\
$0^-$& --- & 12.39  & 13.13  & 12.23 & 12.80\\
$4^-$&18.15&18.15&17.75&17.73&17.79\\
$4^-$&21.41&20.59&19.88&19.45& 18.98\\
\hline\hline
\end{tabular}
\caption{\small Same as in Table \ref{tab:exc12} but for $^{16}$O.
\label{tab:exo16}
}
\end{center}
\end{table}

\begin{figure}[h]
\begin{center}
\includegraphics[scale=0.5, angle=0] {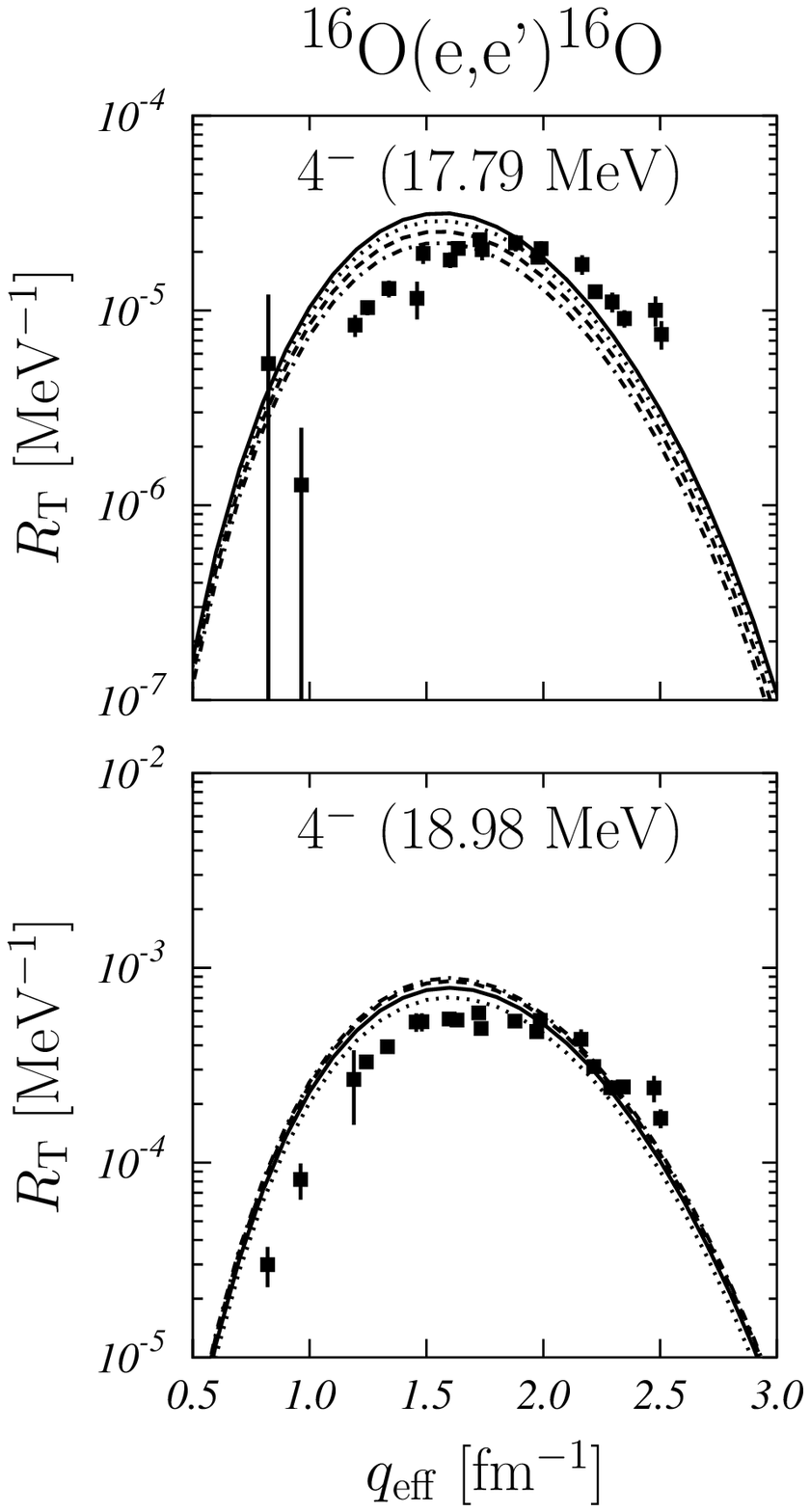} 
\caption{\small Electromagnetic responses of the 4$^-$ states in
  $^{16}$O. The meaning of the lines is the same as in
  Fig. \ref{fig:12C1+}.  The data are taken from \cite{hyd87}.   
  }
\label{fig:16O4-}
\end{center}
\end{figure}

\subsection{The $^{16}$O nucleus}
\label{sec:o16}

The spectrum of the low-lying magnetic states obtained with the four
interactions is presented in Table \ref{tab:exo16}, where it is
compared with the experimental spectrum \cite{led78}.  All the states
have negative parity and, since the $p$-shell is closed for both
protons and neutrons, this indicates that they are dominated by ph
transitions involving neighbor shells. The order of the various
states is reproduced by our calculations. The magnetic state with
lowest energy is a $2^-$ state, as in the experimental spectrum, but
the calculated energy eigenvalues  overestimate the experimental value
of about 30\%, independently of the interaction used.  This result is
contrary to our expectations, because this state is dominated by the
$[(1d_{5/2}) (1p_{1/2})^{-1}]$ bound proton and neutron transitions,
and therefore it should be well described by our approach.  
We have found \cite{don08t} a remarkable
disagreement with the experimental data \cite{hyd87} also for the
transition density. These facts
indicate the presence, in this $2^-$ state, of effects beyond the
description capability of our RPA model.

The subsequent states in our spectrum are two $0^-$ states which can be
identified in the experimental spectrum \cite{led78}. They are
dominated by the $[(2s_{1/2}) (1p_{1/2})^{-1}]$ proton and neutron
transitions. The effect of the tensor term of the interaction on the
energy values is not negligible. Since these states are not excited by
electromagnetic probes, at least in the one-photon exchange picture, we
have calculated neutrino and antineutrino cross sections \cite{don08t}
and we have found large sensitivity to the tensor force.  This point
deserves a more detailed investigation, for example by calculating the
excitations induced by hadronic probes.

The $4^-$ states, dominated by $[(1d_{5/2})(1p_{3/2})^{-1}]$ protons
and neutrons ph excitations, form an isospin doublet. Also in this
case the energy of the IS state is lower than that of the IV one,
in agreement with the experimental data. The IS energy
eigenvalues are almost insensitive to the presence of tensor terms;
they are however rather sensitive to the use of finite-range
interactions. In the IV case, both tensor terms and finite-range
affect the energy value.  The electromagnetic responses for the two
states are shown in Fig.  \ref{fig:16O4-} and compared with the
experimental data of Ref.  \cite{hyd87}. In both cases the position of
the maximum of our calculations is slightly lower than the
experimental one. The IS state shows some sensitivity to the
residual interaction. The inclusion of the tensor term and of the
finite-range contributes to lower the response and this slightly
improves the comparison with the data.  The IV response is less
sensitive to the changes of the interaction.  We obtain a general good
agreement with the data.

\begin{table}[h]
\begin{center}
\begin{tabular}{cccccc}
\hline\hline
\multicolumn{6}{c}{$^{40}$Ca}\\
\hline\hline
$J^\pi$&LM&LMtt&FR&FRtt&exp \\
\hline
$4^-$ &  6.86 &  6.88  &  6.78 &  6.80 &  5.61\\
$2^-$& 7.21 & 7.20& 6.91& 6.90&7.53\\
$4^-$ &  7.52  &  7.59 &  7.42 & 7.47  &  7.66\\
$2^-$&   8.90&    8.44 &     8.76   & 8.58& 8.42\\
\hline\hline
\end{tabular}
\caption{\small Same as in Table \ref{tab:exc12} but for $^{40}$Ca.
\label{tab:exca40}
}
\end{center}
\end{table}

\subsection{The $^{40}$Ca nucleus}
\label{sec:ca40}
The spectrum of the magnetic states of the $^{40}$Ca nucleus is given
in Table \ref{tab:exca40}.  The global closure of the $s$-$d$ shell,
for both protons and neutrons, implies that the 
low-energy spectrum is composed only by negative parity states. Our
RPA calculations reproduce the correct sequence of the states,
independently of the interaction used. The energy eigenvalues do not
show large sensitivity to the choice of the interaction.  We
overestimate the energy of the first $4^-$ state, while the 
energies of the other states are better reproduced.

\begin{figure}[ht]
\begin{center}
\includegraphics[scale=0.5, angle=0] {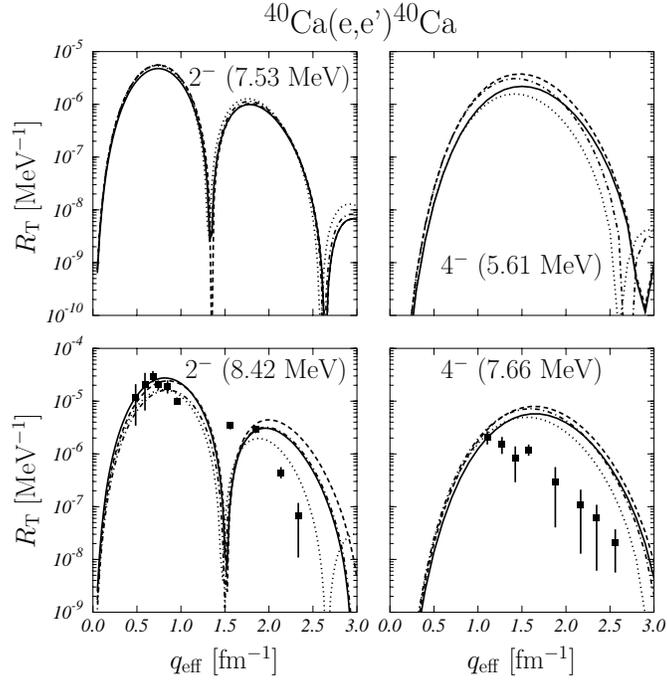} 
\caption{\small Electromagnetic responses of the 2$^-$ and 
  4$^-$ states in $^{40}$Ca. The meaning of the lines is the same as in
  Fig. \ref{fig:12C1+}. The data are taken from \cite{wil87}.
  }
\label{fig:40Ca24-}
\end{center}
\end{figure}

The response functions of the $2^-$ and $4^-$ states are shown in Fig.
\ref{fig:40Ca24-} and compared with the available experimental
data \cite{wil87}. The response of the lowest $2^-$ state is almost
insensitive to the choice of the residual interaction.  The
electromagnetic response of the other $2^-$ state shows larger
sensitivity to the interaction used in the RPA calculation. The shapes
of the responses are strongly modified by the finite-range and
especially
by the tensor term. The latter lowers the value of the first maximum
and reduces the width of the second peak of the response. The data are not  
accurate enough to allow a selection among the various results. 

The response of the lowest $4^-$ state indicates that the presence of
the finite-range increases the peak value, while the tensor term
reduces it. The same effect is present also in the response of the
other $4^-$ state. In this case experimental
data are available for comparison \cite{wil87} and we see that there 
is no similarity in size and shape between our
results and the data.

\subsection{The $^{48}$Ca nucleus}
\label{sec:ca48}

In Table \ref{tab:exca48} we present the low-energy magnetic spectrum
of $^{48}$Ca., obtained  with the same residual
interactions used for  $^{40}$Ca. The $^{48}$Ca
spectrum contains both negative and positive parity states, the latter
being dominated by single particle excitations of the $1f_{7/2}$ neutron
hole. Globally, we obtain a reasonable agreement between the measured
and calculated energies, but the correct sequence of the excited
states is not exactly reproduced. In each calculation we obtain a
6$^-$ state whose energy is larger than that of the 1$^+$ state,
while experimentally the opposite occurs.  This disagreement is due
to the overestimation of the 6$^-$ state energy by about 2.5 MeV. The
energy eigenvalues presented in Table \ref{tab:exca48} do not show
particular sensitivity to the different interactions used in the RPA
calculations

\begin{table}[hb]
\begin{center}
\begin{tabular}{cccccc}
\hline\hline
\multicolumn{6}{c}{$^{48}$Ca} \\
\hline\hline
$J^\pi$&LM&LMtt&FR&FRtt&exp \\
\hline
$3^+$ & 5.03  &  4.99 &   4.96 &   4.94 &   4.61\\
$5^+$ &5.26 &  5.16 & 5.04 & 4.98 &5.15\\
$4^-$ & 6.44 & 6.41 & 6.36 &6.35 &6.10\\
$2^-$ & 7.53 & 7.06 &  7.30 & 7.10 & 6.89\\
$6^-$&11.29 & 11.31 &11.01& 10.99& 8.56 \\
$1^+$&  9.65 &    9.38 &  9.68 &  9.50 &    10.23\\
\hline\hline
\end{tabular}
\caption{\small Same as in Table \ref{tab:exc12} but for $^{48}$Ca.
\label{tab:exca48}
}
\end{center}
\end{table}

\begin{figure}[ht]
\begin{center}
\includegraphics[scale=0.5, angle=0] {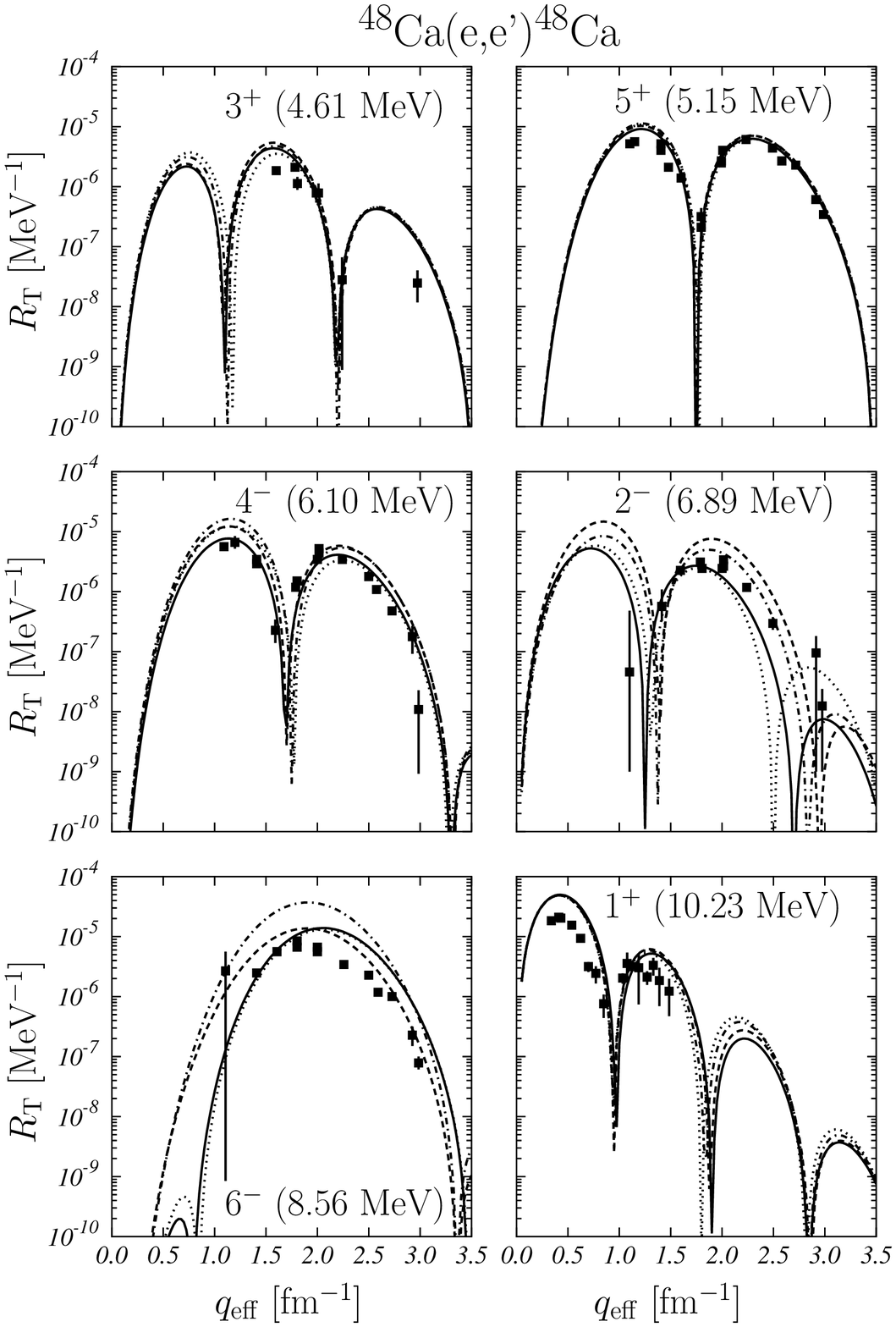} 
\caption{\small Electromagnetic responses of magnetic states
  in $^{48}$Ca. The meaning of the lines is the same as in
  Fig. \ref{fig:12C1+}. The data are taken from \cite{wis85,ste83}.
  }
\label{fig:48Ca}
\end{center}
\end{figure}

The study of the electromagnetic responses of
these states is more interesting, as
 shown in Fig. \ref{fig:48Ca} where the theoretical curves are
compared with the
available experimental data \cite{wis85,ste83}. Large effects of
the choice of the residual interaction are present for  
The 2$^-$,
6$^-$ states look very sensitive to the choice of the residual
interaction, which produces some differences also in the responses
relative to the  4$^-$, $1^+$ and $3^+$ states.  

As already
mentioned, the positive parity states are dominated by the excitation of
the $1f_{7/2}$ neutron hole, therefore the residual interaction plays
a minor role and the results are rather similar to those of the MF.
Effects of the use of different interactions are present 
only at the third maximum of the $1^+$ responses,
and at the first two maxima for the
$3^+$ state. The experimental data of the 3$^+$ and 5$^+$ are rather
well reproduced. The same does not occur with the $1^+$ state where
other mechanisms beyond RPA (second order core polarization, tensor
correlations and $\Delta$ excitations) must be taken into account to
obtain a good agreement between theory and experiment \cite{ama91}.

For the negative parity states we cannot find any
common trend related to the inclusion of the various
ingredients of the interactions. For example, the tensor term
increases the responses of the 4$^-$ and 6$^-$ states, while it lowers
that of the 2$^-$ state. The 2$^-$ and 4$^-$ experimental responses
are rather well reproduced.  Similar results have been obtained in
Ref. \cite{ama92b} where the J\"ulich-Stony Brook interaction
\cite{spe80} with the tensor terms reduced by $\sim$30-60\%  has been
used. 
We have encountered problems in the description of the 6$^-$
response. On the other hand, we have already pointed out the
difficulties found in describing the excitation energy of this state.

\subsection{The $^{208}$Pb nucleus}
\label{sec:pb208}

The energies of the low-lying magnetic states of $^{208}$Pb are
presented in Table \ref{tab:expb208} and compared with the
experimental ones.  The sequence of the states is quite well
reproduced. The are some exceptions, but these occur 
with energy differences of the order of few tens of keV, an
energy resolution smaller than the accuracy we assign to our
results.  The global picture emerging from  Table
\ref{tab:expb208} is that the various interactions produce small
differences in the energy eigenvalues.

\begin{table}[ht]
\begin{center}
\begin{tabular}{cccccc}
\hline\hline
\multicolumn{6}{c}{$^{208}$Pb}\\
\hline\hline
$J^\pi$&LM&LMtt&FR&FRtt&exp \\
\hline
$4^-$ & 3.52 & 3.50 & 3.50 &3.49  & 3.48 \\
$6^-$ & 4.04 & 4.04 & 4.03 & 4.03 & 3.92\\
$2^-$ & 4.30 & 4.21 & 4.23 & 4.20 &4.23\\
$9^+$ & 5.14 & 5.11 & 5.11 & 5.09 & 5.01 \\
$9^+$ & 5.45 & 5.45 & 5.44 & 5.44 & 5.26 \\
$0^-$ & 5.64 & 5.38 & 5.54 & 5.42 &5.28\\
$11^+$ & 5.25 & 5.20 & 5.14 & 5.12 & 5.29 \\
$1^+$ & 5.92 & 5.89 &5.72 &5.70 &5.85\\
$11^+$ & 5.88 & 5.89 & 5.86 & 5.87 & 5.86 \\
$10^-$ & 6.64 & 6.56 & 6.57 & 6.53 & 6.28 \\
$12^-$ &6.66  & 6.61 & 6.57 &6.54  &6.43\\
$14^-$ &6.99 & 6.84 &6.66 &6.59 &6.74\\
$10^-$ & 7.44 & 7.22 & 7.32 & 7.23 & 6.88 \\
$12^-$ &7.72 &7.55 &7.41 &7.32  &7.08\\
$1^+$ & 7.38 & 6.77 &7.64&7.48& 7.30\\
\hline\hline
\end{tabular}
\caption{\small Same as in Table \ref{tab:exc12} but for $^{208}$Pb.
\label{tab:expb208}
}
\end{center}
\end{table}

\begin{figure}[hb]
\begin{center}
\includegraphics[scale=0.5, angle=0] {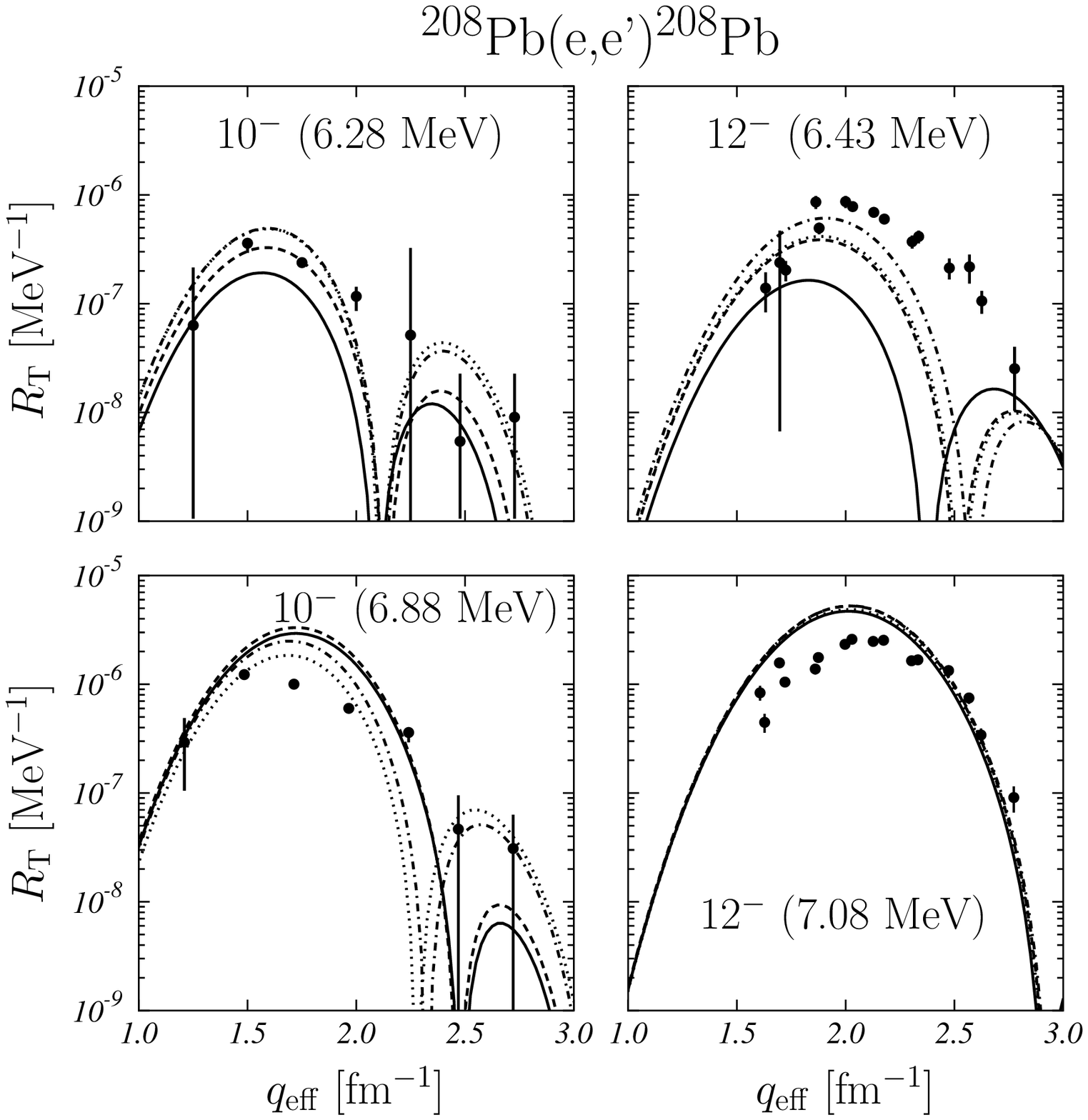} 
\caption{\small Electromagnetic responses of the 10$^-$ and 12$^-$
  states in $^{208}$Pb. The meaning of the lines is the same as in
  Fig. \ref{fig:12C1+}. The data are taken from \cite{lic79}.
 }
\label{fig:208Pb1012}
\end{center}
\end{figure}

The investigation of the electromagnetic responses provides more
information.  We start our discussion with the 12$^-$ responses which
have been studied quite often in the past \cite{kre80,lal88,co90}
because of their, apparently, simple ph structure.  They are, in fact,
mainly composed by two ph pairs, the proton
$[(1i_{13/2})(1h_{11/2})^{-1}]$ and neutron
$[(1j_{15/2})(1i_{13/2})^{-1}]$ transitions.  The lower 12$^-$ state,
experimentally found at 6.43 MeV, is neutron dominated, while the
state at higher energy, 7.08 MeV, is dominated by the proton
transition. Our calculations produce the correct order of the
states, and the RPA energies agree well with the experimental values,
especially the lower one.  We must recall, however, that this state
is one of the states used to set the values of the interaction
parameters.  
The calculated energies of the higher state overestimate the experimental
value, but the discrepancies are below 10\%.  The electromagnetic
responses are shown in the right panels of Fig. \ref{fig:208Pb1012}
and are compared with the data of Ref.  \cite{lic79}. The responses
relative to the higher state, lower panel, show a reasonable agreement
with the data and they are almost insensitive to the choice of the
residual interaction.  On the contrary, the responses of the neutronic
state, upper panel, are extremely sensitive to the inclusion  of both
finite-range and tensor terms in the interaction. These effects
improve the agreement with the data, but the calculated curves still
underestimate the measured response.  The disagreement could be
reduced by increasing the magnitude of the tensor part of the
interaction. We have found, however, that this would produce 
a general worsening
of the magnetic spectrum of $^{208}$Pb, and also of the other nuclei
we have considered. For example a too strong tensor interaction
could invert the sequence of the IS and IV $1^+$ states.

\begin{figure}[ht]
\begin{center}
\includegraphics[scale=0.5, angle=0] {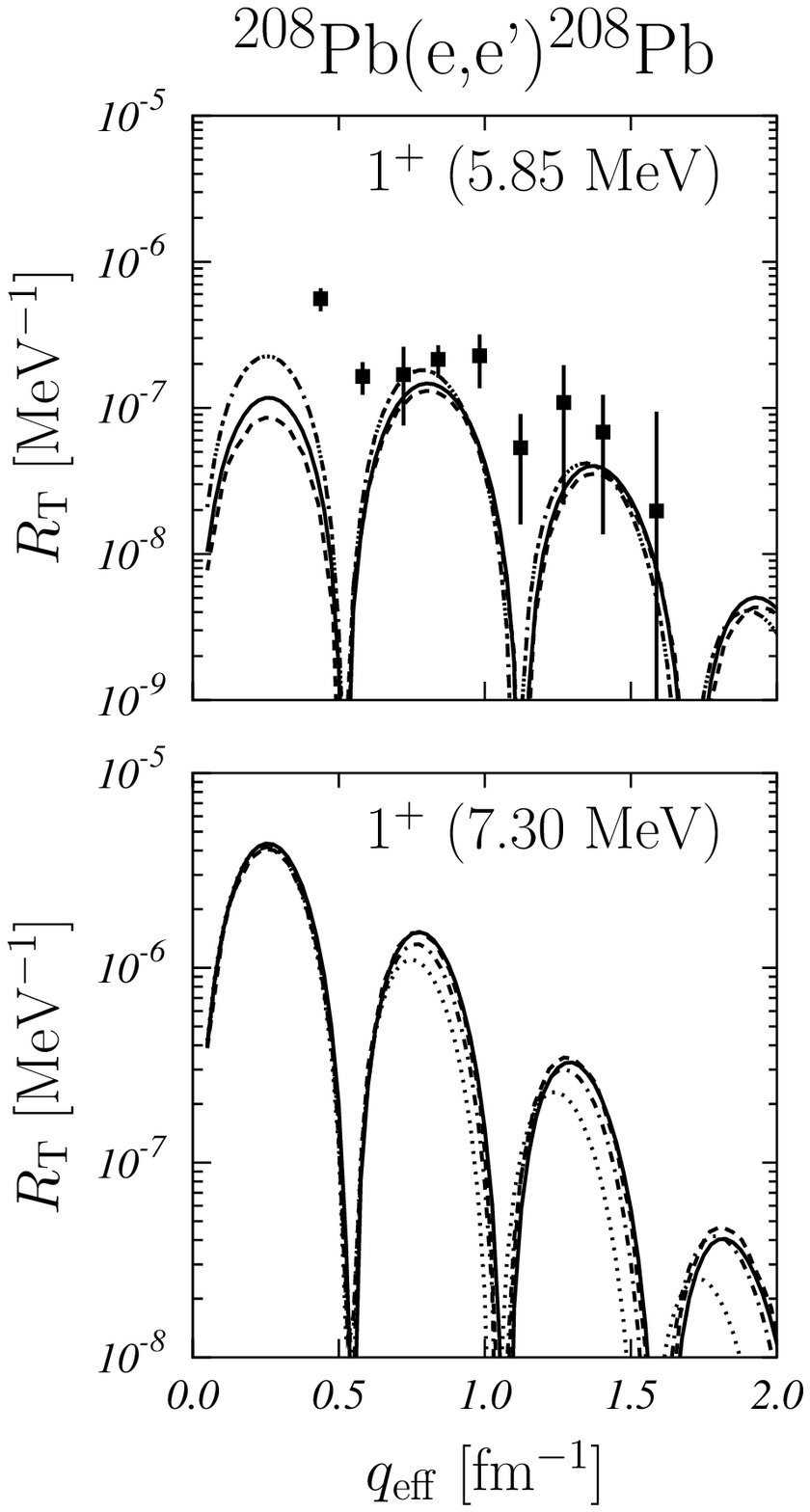} 
\caption{\small Electromagnetic responses of the 1$^+$ 
  states in $^{208}$Pb. The meaning of the lines is the same as in
  Fig. \ref{fig:12C1+}. The data are taken from \cite{mul85}.
 }
\label{fig:208Pb1}
\end{center}
\end{figure}

\begin{figure}[ht]
\begin{center}
\includegraphics[scale=0.5, angle=0] {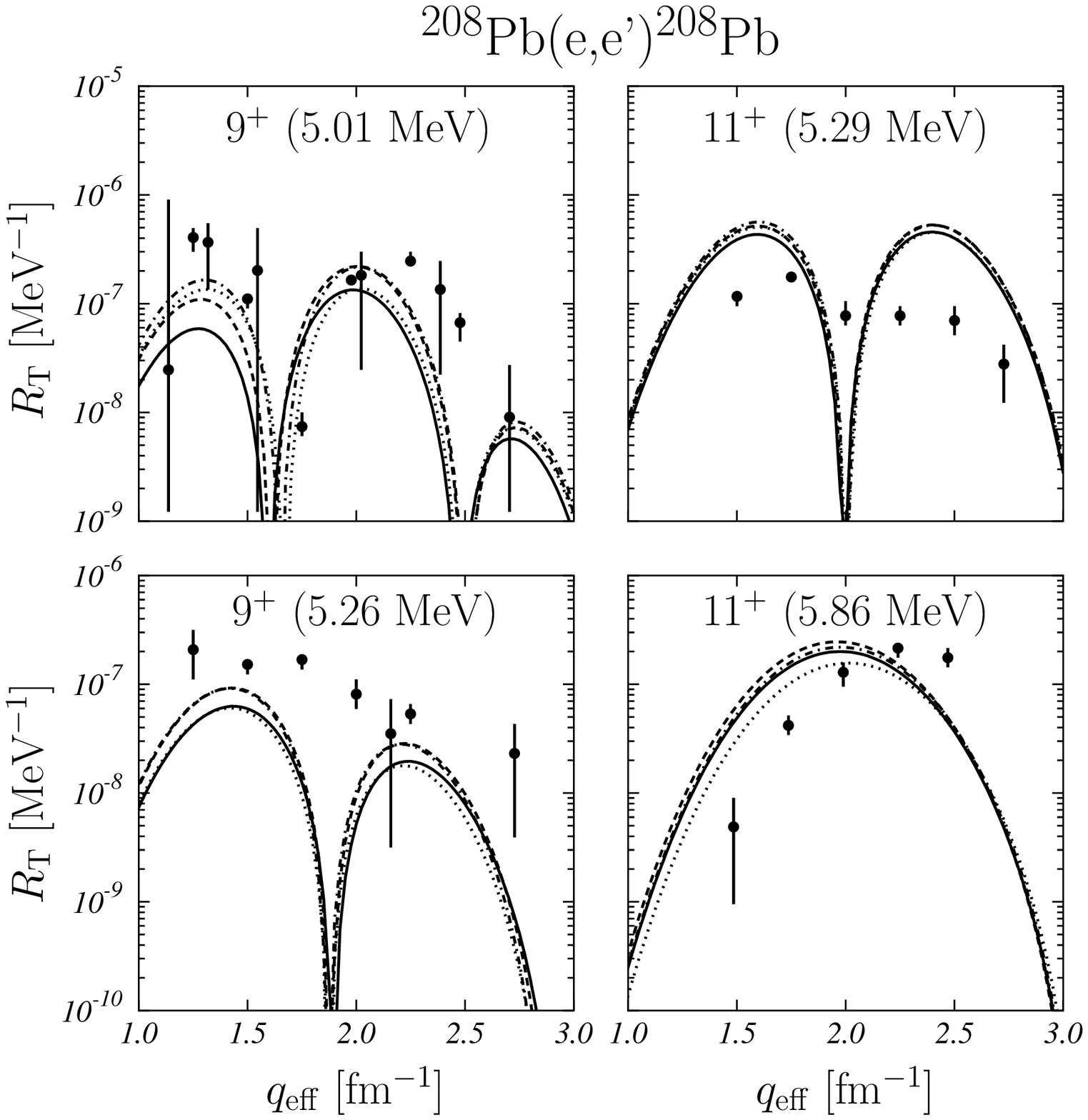} 
\caption{\small Electromagnetic responses of the 9$^+$ and 11$^+$
  states in $^{208}$Pb. The meaning of the lines is the same as in
  Fig. \ref{fig:12C1+}. The data are taken from \cite{lic79}.
 }
\label{fig:208Pb911}
\end{center}
\end{figure}

In the left panels of Fig. \ref{fig:208Pb1012} we present the
responses of the $10^-$ states, which show some sensitivity to the
the tensor part of the residual interaction.  All
interactions can reproduce the magnitude of the responses, but only
the inclusion of the tensor terms allows a good description of the
second peak in both $10^-$ states.  Improvements in the precision of
experimental data around $q=2.5$ fm$^{-1}$ would  thus be particularly
important to study the tensor component of the residual interaction.

The energies of the $1^+$ states are reproduced rather well with all
the residual interactions, both for the isoscalar state at $5.85$ MeV and
for the isovector state at $7.30$ MeV. We should point out that the IV
state is so fragmented that this energy value is an estimate based on
an accurate analysis of the photon scattering data \cite{las85}.  The
electromagnetic responses are plotted in Fig. \ref{fig:208Pb1}, and
for the IS state (upper panel) we compare them with the data
\cite{mul85}.  For this state, all the curves reproduce the
$q$-dependence of the data except at low $q$, where the
theoretical responses are well below the data. Unfortunately there are
no data in the region $q<$ 0.5 fm$^{-1}$ where the effects of the
different interactions are larger.  
In the IV case (lower panel) the differences between the
various results appear at large $q$ values.  This is, however, a
theoretical speculation, because, as we have already said,
experimentally the IV state is extremely fragmented, and cannot be
described within our RPA approach. In the \pb nucleus, pairing effects
are negligible, and we think that this fragmentation can be described
only by considering elementary excitations beyond 1p-1h.  

The $9^+$, $11^+$ and $14^-$ states are dominated by a single
particle-hole excitation, with the exception of the lower energy $9^+$
state, where a small contribution of the proton
$[(2f_{7/2})(1h_{11/2})^{-1}]$ transition is present besides the
dominant neutron $[(2g_{9/2})(1i_{13/2})^{-1}]$ one.  For this state,
the calculated transverse responses, presented in the upper left panel
of Fig. \ref{fig:208Pb911}, show three peaks, and this behavior is
compatible with the data.  On the other hand, the position of the
experimental points of the other $9^+$ state, shown in the lower left
panel of the same figure, is very different from the shape of the
theoretical responses, which exhibit some dependence on the residual
interaction.  Analogous problems are found for the $11^+$ states,
whose responses are plotted in the right panels of Fig.
\ref{fig:208Pb911}.

\begin{figure}[hb]
\begin{center}
\includegraphics[scale=0.5, angle=0] {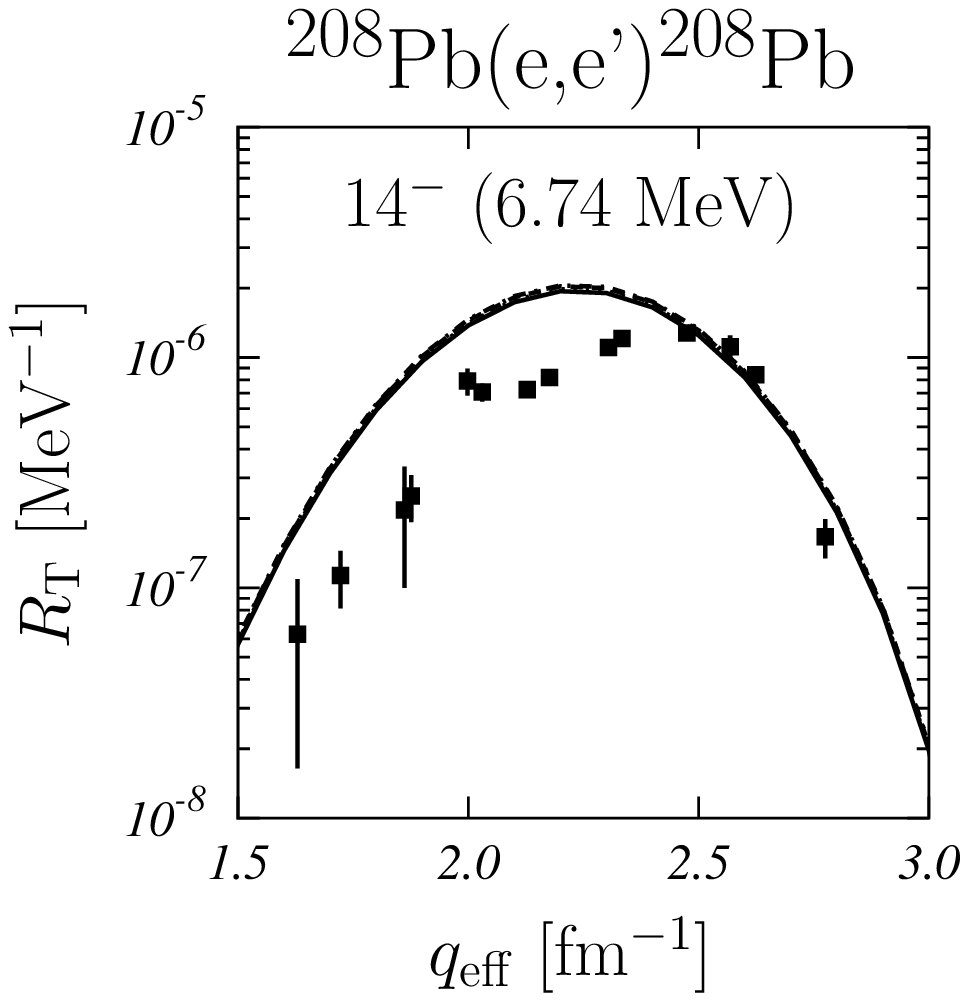} 
\caption{\small Electromagnetic responses of the 14$^-$ 
  states in $^{208}$Pb. The meaning of the lines is the same as in
  Fig. \ref{fig:12C1+}. The data are taken from \cite{lic79}.
 }
\label{fig:208Pb14}
\end{center}
\end{figure}

To complete our survey we show in Fig. \ref{fig:208Pb14} the
electromagnetic response of the $14^-$ state. Its nature of almost
pure ph transition is evident because there is no dependence on the
residual interaction, as pointed out in the literature (see for
example \cite{hin92}).

\section{Results of self-consistent calculations}
\label{sec:HF}

In the previous section we have presented the results of the
phenomenological approach.  We would like to point out that the study
of the full set of magnetic states, together with their
electromagnetic responses, can be used to test the validity of the
effective interactions used in RPA calculations. In order to give an
example of this potentiality, we present here some selected results we
have obtained with the Gogny D1 interaction \cite{gog75,bla77,dec80}.

The complete D1 interaction has been used in computing the HF single
particle energies and wave functions, while we have neglected the
contribution of the spin-orbit term in the RPA calculations. This is a
good approximation if also the contribution of the residual Coulomb
interaction is neglected \cite{sil06}, as we have done. The RPA
results presented in this section have been obtained by considering
both direct and exchange terms of the D1 interaction, in analogy to
what we have done in the Hartree-Fock calculations.  This makes the
connection between the properties of the excitation spectrum and the
various parts of the interaction much more complicated than in the
phenomenological approach, where only direct matrix elements have been
considered, because each interaction term can now contribute to the
other channels through the exchange diagrams.  For example, in the
phenomenological case scalar and isospin channels do not contribute to
the excitation of unnatural parity states, whereas these two channels
produce a contribution to the spin and spin-isospin channels in the
exchange diagrams in the calculations done with the D1 interaction.

To complete the information about our RPA calculations we point out
that we have also included the so-called rearrangement terms, related
to the density dependent part of the interaction. They arise by
considering the effective interaction as the second derivative of the
energy with respect to the single particle density
\cite{rin80}. Quantitatively, we have found the contributions of these
terms to be negligible in all the cases we have investigated.

\begin{figure}[ht]
\begin{center}
\includegraphics[scale=0.5, angle=0] {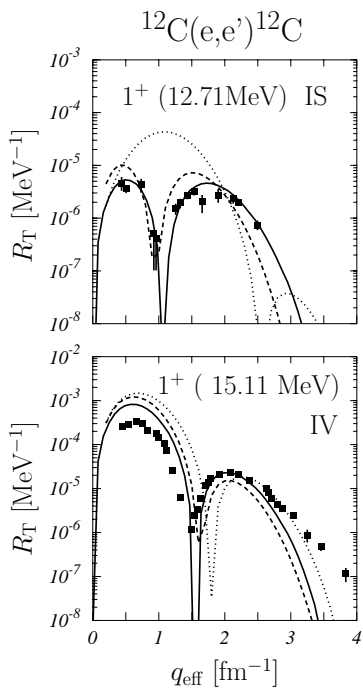} 
\caption{\small Electromagnetic responses of the 1$^+$ isospin doublet 
  in $^{12}$C. The full lines show the results of Fig. \ref{fig:12C1+}
  obtained with the FR interaction. The dotted lines have been
  obtained with the D1 interaction but using the set of single particle wave
  functions and energies used in the phenomenological approach. The
  dashed lines are the results a  self-consistent calculation
  with the D1 interaction. This means that the single particle basis has been
  generated by a Hartree-Fock calculation with the D1 interaction.
  The dotted and dashed curves in the IS panel have been obtained by
  using the RPA amplitudes of the higher energy 1$^+$ solution. The
  lower energy amplitudes have been used to generate the curves 
  shown in the IV panel.     
  }
\label{fig:12CD11+}
\end{center}
\end{figure}

\begin{figure}[ht]
\begin{center}
\includegraphics[scale=0.45, angle=90] {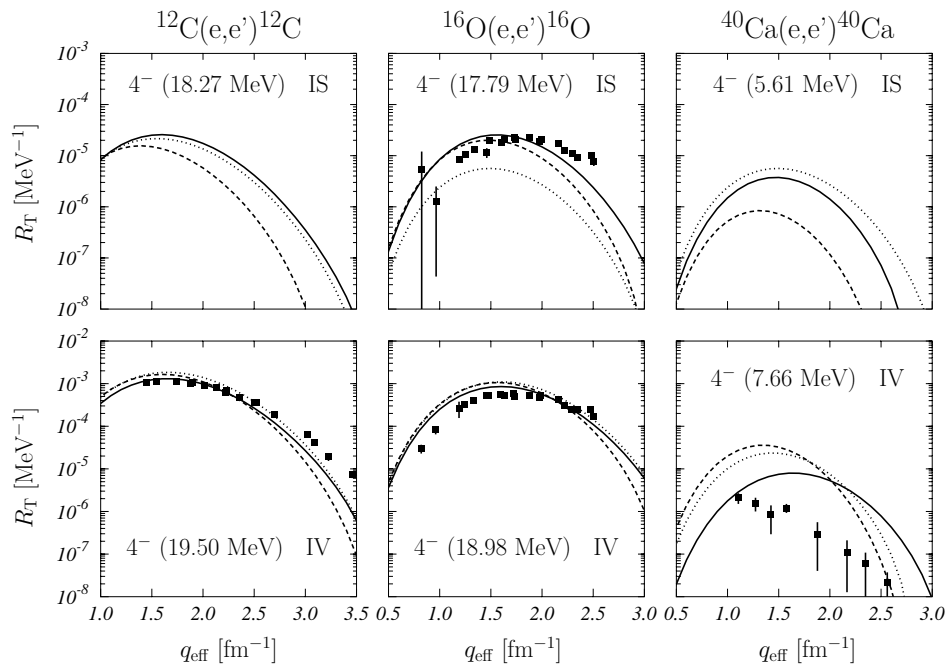}  
\caption{\small Electromagnetic responses of some 4$^-$
 isospin doublets. The meaning of the lines is the same as in 
 Fig. \ref{fig:12CD11+}. Also in this case the low energy D1 responses
 are plotted together with the high energy phenomenological responses,
 and vice-versa.  
  }
\label{fig:COD14-}
\end{center}
\end{figure}

\begin{figure}[ht]
\begin{center}
\includegraphics[scale=0.5, angle=0] {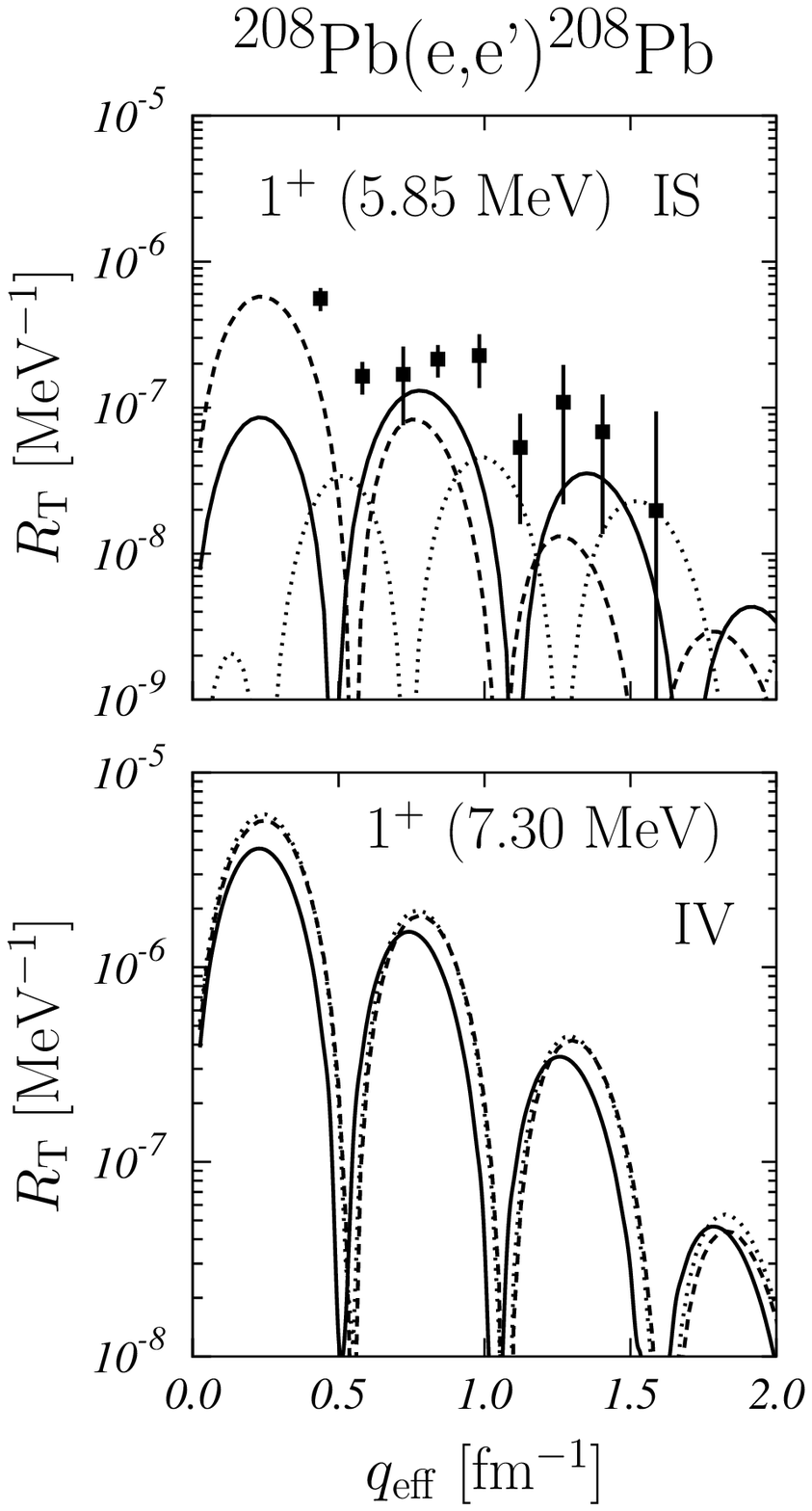}  
\caption{\small The same as  Fig. \ref{fig:12CD11+} for the 1$^+$
  isospin doublet in \pb.
  }
\label{fig:208PbD11+}
\end{center}
\end{figure}

In the following, we present two different types of results.  In the
first case the Gogny D1 interaction is used in the RPA calculations,
but the size of the configuration space and the single particle wave
functions and energies are taken to be the same as in the
phenomenological approach. The results of these calculations are
represented by the dotted lines in Figs.  \ref{fig:12CD11+},
\ref{fig:COD14-} and \ref{fig:208PbD11+}. The second case is fully
self-consistent, i.e. the single particle basis is produced by a
Hartree-Fock calculation with the same interaction used in RPA.  The
size of the configuration space is chosen as described in
Sect. \ref{sec:details}.  The corresponding results are shown in the
figures as dashed lines. The comparison between these two cases allows
us to distinguish between the role played by the single particle basis
and that played by the residual interaction. In the figures, the full
lines show the results obtained in the phenomenological approach by
using the FR interaction.  This interaction has finite-range but it
does not include the tensor terms, therefore, among the four
interactions we have defined in the previous section, it is the most
similar to the D1 interaction.

We start our discussion with the 1$^+$ isospin doublet in $^{12}$C,
whose electromagnetic responses are shown in Fig. \ref{fig:12CD11+}.
The striking result is that when the D1 interaction is used the
position of the IS and IV states is inverted. 
  In the self-consistent calculations with the D1 interaction we
  obtain the lowest 1$^+$ state at 7.72 MeV. In Fig. \ref{fig:12CD11+}
  the response obtained with the RPA amplitudes of this state is
  presented in the lower panel, together with the data and with the
  phenomenological response for the IV state. The self-consistent
  response obtained with the D1 interaction at 10.66 MeV is shown in
  the upper panel of the figure, together with the IS phenomenological
  response.

  We obtain this inversion also in the calculations done with the 
  D1 interaction and the phenomenological single particle wave functions 
  and energies (dotted curves). The response function of the lowest energy
  state, at 3.85 MeV, is shown in the lower panel together with the
  IV data. In the upper panel of the figure we show the response
  obtained by using the RPA amplitude of the state at 8.12 MeV. 
  It is not simple to identify the IS state among those we have
  obtained  in this energy region. We have chosen the state showing
  large values of the RPA amplitude for the 
  $[(1p_{1/2})(1p_{3/2})^{-1}]$ proton and neutron transitions. The
  shape of this response is very different from that of the other
  responses and from the data.

We have found the inversion of the IS and IV partner states in all the
cases we have investigated. Examples are shown in
Fig. \ref{fig:COD14-} for a set of 4$^-$ states 
and in Fig. \ref{fig:208PbD11+} for the 1$^+$ states in \pb. 

The energies, in MeV, of the self-consistent calculations of the 4$^-$
states are: 1.64 and 18.64 for \car, 15.49 and 18.81 for \oxy and 7.59
and 7.83 for \caI.  The comparison with the phenomenological results,
and with the experimental data, is always done by associating the
responses corresponding to the higher energy values to the IS states,
and those corresponding to lower energy to the IV states.  The
energies of the \pb 1$^+$ states are 6.75 MeV (IV) and 9.40 MeV (IS).

We stress that the inversion is obtained in both types of
RPA calculations done with the D1 interaction and therefore it does
not depend on the single particle basis, but it is related to the
characteristics of the interaction itself.
We have repeated our calculations with another Gogny-like force with
different values of the parameters, the D1S interaction \cite{ber91}, and
also in this case we have observed the inversion of the isospin partner
states.

\section{Conclusions}   
\label{sec:conclu}

We have studied the magnetic excitation spectrum of doubly closed
nuclei to investigate the properties of the spin, spin-isospin and
tensor terms of the effective interaction. In a phenomenological
approach, where the single particle basis is obtained by using
Woods-Saxon wells, we have introduced four different
interactions which reproduce  the energy of
specific magnetic excited states in \car, \oxy, \caI, \caII and \pb
with the same  accuracy.
We have  first considered a zero-range interaction having only the four
central channels, and we have then  progressively complicated the structure
of the interaction by adding tensor terms and finite-range.  The RPA
calculations we have done for a large number of magnetic excitations 
indicate
that all four interactions are able to describe with reasonable
accuracy the experimental spectra and, to a lesser extent,
the electromagnetic responses.
We have found, and pointed out, a few cases 
where the role of the finite-range and of the
tensor terms is relevant, for example the neutronic 12$^-$ state of
$^{208}$Pb shown in Fig.  \ref{fig:208Pb1012}. 

In some cases we have found large disagreement between our
calculations and the experimental data, as, for
example, for the 4$^-$ state of $^{40}$Ca at 7.66 MeV, shown in Fig.
\ref{fig:40Ca24-}. In these cases however, the discrepancies between 
calculations and data are more related to the inadequacy of
the RPA description rather than to a bad parametrization of the
interaction.

The validity of our approach has then been tested with the Gogny D1
interaction, for which we have repeated the calculations of the magnetic
excitation of all the states considered in the phenomenological
approach. The calculations have been done
both using  the same single particle basis employed 
in the phenomenological case
and  in a fully self-consistent approach, where the single
particle basis has been generated by a Hartree-Fock calculation. The
striking result we have obtained is that the D1 interaction inverts
the energy sequence of isospin partner excitations, independently of
the single particle basis adopted and for all the nuclei studied. For
a fixed multipolarity the experimental evidence is that the IS
excitation has lower energy than the IV one, while this order is
inverted in the RPA calculations with the D1 interaction.  Nuclear
matter studies of the pairing gap done with the D1 interaction
indicate anomalous behavior in the isospin T=0 channel
\cite{gar99}. The two problems could be related.  In these
circumstances, the role of both the spin-orbit and the residual
Coulomb terms of the interaction, which are neglected in our RPA
calculations, should be investigated in order to control the validity
of the D1-like interactions for this kind of calculations.

Improvements of Gogny-like interactions have recently attracted a
lot of attention \cite{ots06,cha06,ots07}, because self-consistent
calculations have a wider predictive power than phenomenological
approaches.  The description of exotic nuclei, which will be produced
and studied in the future nuclear physics facilities, requires the use
of well grounded self-consistent calculations. We think that the
analysis of the magnetic spectra and of their electromagnetic
properties is an important filter to select the nucleon-nucleon
interactions to be used in effective nuclear theories.

\acknowledgments
This work has been partially supported by the Spanish Ministerio
de Ciencia e Innovaci\'on under contract FPA2008-04688 and by the Junta
de Andaluc\'{\i}a (FQM0220).


\end{document}